# ROSAT PSPC observations of the Seyfert 1 galaxies Ark 564, NGC 985, Kaz 163, Mrk 79 and RX J2256.6+0525

W.N. Brandt,[1] A.C. Fabian,[1] K. Nandra,[1] C.S. Reynolds[1] and W. Brinkmann[2]

[1] *Institute of Astronomy, Madingley Road, Cambridge CB3 0HA*
[2] *Max-Planck-Institut für Extraterrestrische Physik, D-85478, Garching, Germany*

**ABSTRACT**

We present spatial, temporal and spectral analyses of *ROSAT* Position Sensitive Proportional Counter (PSPC) observations of the Seyfert 1 galaxies Ark 564, NGC 985, Kaz 163, Mrk 79 and RX J2256.6 + 0525. Ark 564 is a powerful narrow-line Seyfert 1 with strong Fe II emission. Several similar narrow-line Seyfert 1 galaxies have recently been found to have remarkably steep soft X-ray continua as well as rapid X-ray variability. We find that Ark 564 also has a very steep ($\Gamma > 3$) 0.1–2.5 keV spectrum and varies by $\sim 20$ per cent in 1 500 s. We examine models for Ark 564 in light of both its X-ray and optical characteristics, and suggest a possible connection between the steep X-ray spectra of narrow-line Seyfert 1 galaxies and their narrow lines. NGC 985 has a large soft excess, a warm absorber, or both. The three other Seyferts have systematically steeper spectra than are typically observed for Seyferts in higher energy *Ginga* data, indicating that they also harbour further spectral complexity. Kaz 163 shows $\sim 45$ per cent intensity variability in the 0.1–2.5 keV band, and Mrk 79 shows evidence for variability as well.

## 1 INTRODUCTION

X-ray studies of Seyfert 1 galaxies penetrate into the vigorous central engines of these objects, showing the signature of the fuel that powers them (Mushotzky, Done & Pounds 1993). Recent high signal-to-noise ratio spectra obtained by *Ginga* in the 2–20 keV band have revealed an underlying power law with photon index $\Gamma = 1.95^{+0.05}_{-0.05}$ (Nandra & Pounds 1994) as well as a flattening of the spectrum (the reflected 'hard tail') above 10 keV. Excess absorption is found in about half of Seyfert 1s, and a fluorescent iron K$\alpha$ line is seen in over 90 per cent of them. At least 40 per cent of Seyfert 1s appear to have warm absorbers, which revealed themselves in *Ginga* data through their 8–9 keV ionized iron K edges. In the soft X-ray band a variety of spectral shapes are seen. Differing amounts of photoelectric absorption and reprocessing, line emission, and Compton reflection lead to complex soft X-ray spectra with multiple continua as well as sharper spectral features. Warm absorption by ionized oxygen has been seen by *ROSAT* in several Seyferts (e.g. Nandra et al. 1993). Here we analyse five X-ray Seyfert 1 galaxies, whose soft X-ray properties have until now been studied in only modest detail, as part of an effort to develop a unified taxonomy of X-ray Seyferts. Four of the five objects were selected as bright Seyfert 1 galaxies with potentially complex soft spectra in the *ROSAT* all-sky survey.

Arakelian 564 (MCG +05 − 53 − 012) is classified as a narrow-line Seyfert 1 (Osterbrock & Shuder 1982; see Osterbrock & Pogge 1985 for the criteria used to define the narrow-line Seyfert 1 class of objects) and has relatively strong [Ca II] emission and strong Fe II emission lines (van Gronigen 1993). Recent *ROSAT* observations of similar narrow-line Seyfert 1s (e.g. IRAS 13224 − 3809, IZw 1, 5C3.100, RE 1034+393, Mrk 478, Mrk 766, Mrk 1044) have shown extremely steep soft X-ray continua and fast X-ray variability to be features often seen in these galaxies (Boller et al. 1993; Boller, Brandt & Lipari, in preparation; Gondhalekar et al. 1994). IRAS 13224 − 3809, for example, has a photon index $> 4$, a 0.1–2.4 keV luminosity of $3 \times 10^{44}$ erg s$^{-1}$ and a doubling time of 800 s, giving it one of the largest currently known compactness parameters for a Seyfert of $\approx 170$ (Boller et al. 1993; Fabian 1992). The large number of narrow-line Seyfert 1 galaxies that appear in the *Einstein* IPC ultra-soft survey corroborates the steep spectra seen by *ROSAT* (see table 5 of Puchnarewicz et al. 1992). Ark 564 was one of the peculiar outlying Seyferts in the Walter & Fink (1993) *ROSAT* all-sky survey correlation between spectral slope and ultraviolet (1375 Å) to 2-keV flux (as was Mrk 766, another narrow-line Seyfert 1). Walter & Fink (1993) argue that it has a peculiarly weak ul-traviolet continuum because a large part of the narrow-line region is covered by obscuring dust. Ark 564 has a narrow-line Balmer decrement of 4.4 (van Groningen 1993; Walter & Fink 1993).

NGC 985 (Mrk 1048, VV 285) is a morphologically peculiar Seyfert 1. One-hundred and eight years ago, F.D. Leavenworth noted its strong quasi-stellar nucleus (Stone 1886). NGC 985 has a giant ($\sim 40$ kpc) ring-shaped zone of strong infrared emission, which Rodríguez Espinosa & Stanga (1990) suggest arises from hot gas being photoionized by OB stars formed in an intense episode of star formation. Stanga, Rodríguez Espinosa & Mannucci (1991) have found a nuclear double-peaked He I line, and also demonstrate that the H$\alpha$ and H$\beta$ lines are asymmetric. They are able to fit the line profiles consistently with an optically thick Keplerian accretion disc model, but point out that this is not the only possible interpretation of the lines. NGC 985 is a strong X-ray source, having a 0.2–4.0 keV luminosity of $1.4 \times 10^{44}$ erg s$^{-1}$ in the observation of Kruper, Urry & Canizares (1990).

The Seyfert 1 galaxy Kazarian 163 (VII Zw 742) is in the process of collision with an object to the north resembling an elliptical galaxy in shape and colour (Kriss & Canizares 1982; Hutchings & Hickson 1988). It lies in an irregular blue nebulosity and its nuclear spectrum has relatively weak [O III] lines and some Fe II blends (Hutchings & Hickson 1988). The ratio of its permitted and forbidden lines falls away from the nucleus, but then rises again in outer parts of the nebulosity. Kaz 163 has been observed in the *ROSAT* all-sky survey (Walter & Fink 1993) and in the *Einstein* EMSS (Kriss & Canizares 1982; Stocke et al. 1991), but no spectral features or variability were observed.

Markarian 79 (MCG $+08 - 14 - 033$) is a barred Seyfert 1.2 and is also an extended ($\approx 2$ kpc) three-component linear radio source (Ulvestad & Wilson 1984, Mazzarella & Boroson 1993). In addition to its broad permitted lines, Mrk 79 has strong Fe II lines as well as the stronger forbidden lines such as [O II], [O III] and [Ne III] (Oke & Lauer 1979). Mrk 79 was studied by *EXOSAT*, but noise in the *EXOSAT* data resulted in a poor quality spectral fit (Turner & Pounds 1989).

RX J2256.6 + 0525 (NAP 3) lies 5 arcmin on the sky away from the $z = 0.196$ X-ray cluster Abell 2507. It was recently discovered in optical spectroscopy by Crawford et al. (1994), and has significantly contaminated X-ray studies of Abell 2507 made using *Ginga* and *HEAO-1* (Johnson et al. 1983; Arnaud et al. 1991). Fig. 1 shows the optical spectrum of RX J2256.6 + 0525, which lies at $z = 0.066$. The H$\beta$ line of RX J2256.6 + 0525 has easily visible broad- and narrow-line components, and RX J2256.6+0525 is best classified as a Seyfert 1.5. The H$\beta$ broad-line FWHM is $\approx 3\,000$ km s$^{-1}$ and the [O III] narrow-line FWHM is $\approx 800$ km s$^{-1}$.

A value of the Hubble constant of $H_0 = 50$ km s$^{-1}$ Mpc$^{-1}$ and a cosmological deceleration parameter of $q_0 = \frac{1}{2}$ have been assumed throughout.

## 2   OBSERVATIONS AND DATA REDUCTION

*ROSAT* PSPC (Trümper 1983; Pfeffermann et al. 1987) observations were made of the five Seyferts on the dates shown in Table 1. Also shown in Table 1 are redshifts, exposure times, raw background-subtracted counts, absorbed fluxes, unabsorbed fluxes and Galactic $N_H$ values. The *ROSAT* observations were performed in the standard 'wobble' mode; to avoid accidental shadowing of sources by the coarse wire grid which forms part of the PSPC entrance window support structure, *ROSAT* performs a slow dithering motion diagonal to the detector axes with a period of $\sim 400$ s and an amplitude of 3 arcmin.

The PSPC detector at the focus of the *ROSAT* X-ray telescope has a bandpass of $\sim 0.1$–2.5 keV over a $2^{\circ}$-diameter field of view. Its energy resolution is $\Delta E/E = 0.43(E/0.93)^{-0.5}$ (FWHM), where $E$ is the photon energy measured in keV. The spatial resolution for the full bandpass is $\approx 25$ arcsec (FWHM) on-axis (Hasinger et al. 1992). The PSPC background count rate is very low, and is dominated by the cosmic ray induced particle background, scattered solar X-rays and the diffuse X-ray background. The rejection efficiency for the particle background is excellent, and the residual count rate due to cosmic rays is only $\sim 5 \times 10^{-6}$ count s$^{-1}$ arcmin$^{-2}$ keV$^{-1}$ (Snowden et al 1992).

Reduction and analysis of the PSPC image data was performed with the Starlink ASTERIX X-ray data processing system.

## 3   ANALYSES

### 3.1   Spatial analyses

The 0.1–2.5 keV spatial profiles of all targets are consistent with those of point sources convolved with the *ROSAT* XRT and PSPC spatial responses (Hasinger et al. 1992). None of the targets was detected with the *ROSAT* Wide Field Camera. All targets lie in the centres of their fields except for Kaz 163 and RX J2256.6+0525. Kaz 163 lies 10.6 arcmin away from the field centred on Mrk 507, and RX J2256.6 + 0525 lies 5.5 arcmin from the field centred on Abell 2507.

We have extracted the source counts for all objects from circular source cells chosen to be large enough to ensure that all of the source counts are included, given the electronic 'ghost imaging' which widens the point spread function below $\sim 0.3$ keV (Hasinger et al. 1992). Background counts were subtracted from the source cells using large circular source-free background cells. Corrections were included for PSPC dead time, vignetting and shadowing by the coarse mesh window support. We use these data in the temporal and spectral analyses below.

### 3.2   Temporal analyses

Count rates for X-ray sources can only be reliably given as averages over the *ROSAT* wobble period of $\sim 400$ s, as the path of the source in the detector should repeat with this period. The wobble path can be geometrically distorted by digitization uncertainties in the star sensor read-out, and this can lead to apparent variations in count rate over different wobble periods. However, such effects are generally small and the accuracy of flux determinations in wobble mode are good to within $\approx 4$ per cent (Brinkmann et al. 1994). We have used 400-s binning for all temporal analysis work. In addition, we have conservatively disregarded data from the first and last 20 s of each observation interval so as to avoid potential detector voltage and satellite drift problems.

Kaz 163 shows $\sim 45$ per cent X-ray flux variability, and its 0.1–2.5 keV light curve (corrected for detector dead

**Figure 1.** The optical spectrum of the $z = 0.066$ X-ray Seyfert 1.5, RX J2256.6 + 0525. The spectrum has not been de-reddened.

**Table 1.** Observational parameters.

| Seyfert Galaxy Name | $z$ | Exposure Time (s) | Observation Date (Mission Phase) | Raw bckgnd. subtracted counts | 0.2–2.0 keV Absorbed Flux (ergs cm$^{-2}$ s$^{-1}$) | 0.2–2.0 keV Unabsorbed Flux (ergs cm$^{-2}$ s$^{-1}$) | Galactic $N_{\rm H}$ /($10^{20}$ cm$^{-2}$) |
|---|---|---|---|---|---|---|---|
| Ark 564 | 0.024 | 4 316 | 30 Nov 93 (AO-4) | 13 080 | $3.44 \times 10^{-11}$ | $9.79 \times 10^{-11}$ | $6.1^{+0.3}_{-0.3}$ |
| NGC 985 | 0.043 | 6 194 | 09 Aug 93 (AO-4) | 7 860 | $1.32 \times 10^{-11}$ | $2.07 \times 10^{-11}$ | $3.0^{+0.2}_{-0.2}$ |
| Kaz 163 | 0.063 | 24 754 | 08 Aug 93 (AO-4) | 6 710 | $0.28 \times 10^{-11}$ | $0.48 \times 10^{-11}$ | $4.3^{+0.2}_{-0.2}$ |
| Mrk 79 | 0.022 | 2 694 | 24 Sep 93 (AO-4) | 5 010 | $2.31 \times 10^{-11}$ | $5.31 \times 10^{-11}$ | $5.9^{+0.3}_{-0.3}$ |
| RX J2256.6+0525 | 0.066 | 5 318 | 25 Nov 91 (AO-2) | 1 540 | $0.35 \times 10^{-11}$ | $0.81 \times 10^{-11}$ | $5.5^{+0.2}_{-0.2}$ |

Fluxes are computed using the preferred models from the text. The Galactic column density for Mrk 79 is taken from Elvis, Lockman & Wilkes (1989). The Galactic column densities for all other sources are taken from Stark et al. (1992).

time, vignetting and wire shadowing) is shown in Fig. 2. The hardness ratio for Kaz 163 does not change significantly when its flux changes. Ark 564 has three observation intervals. During the first, the light curve of Ark 564 smoothly decreases by $\sim 20$ per cent over 1 500 s from $3.8 \pm 0.1$ count s$^{-1}$ to $3.0 \pm 0.1$ count s$^{-1}$. By the second observation interval, $\sim 80\,000$ seconds later, Ark 564 registers $3.4 \pm 0.1$ count s$^{-1}$. During the third observation interval, $\sim 80\,000$ seconds after the second, Ark 564 registers $4.0 \pm 0.1$ count s$^{-1}$. Ark 564 showed large amounts of variability during the *HEAO-1* all-sky survey as well. Its 1–20 keV flux, as measured by *HEAO-1*, was 1.01 mC, 0.62 mC, and $< 0.3$ mC on 16 Dec 1977, 11 June 1978, and 16 Dec 1978 (R. Mushotzky, private communication). Due to its short exposure time, Mrk 79 has only two *ROSAT* observation intervals, separated by $3.5 \times 10^5$ s. During the first, Mrk 79 gives $2.2 \pm 0.1$ count s$^{-1}$, and during the second it gives $3.2 \pm 0.1$ count s$^{-1}$.

At present we cannot state with confidence whether or not we are seeing real intensity variations of our sources on time-scales shorter than 400 s or whether the observed fluctuations are the result of the irregular wobble motion. Such analysis will have to await further understanding of the wobble's effect on short-term source variability.

We have extensively searched for instrumental effects that might produce the variability seen in these objects, and can exclude them. We have examined the light curves of other serendipitous sources in the *ROSAT* fields of view, and they show no signs of correlated variability with the objects discussed above.

### 3.3 Spectral analyses

Counts from the corrected circular source cells were extracted into 256-channel, pulse-invariant spectra. We ignored channels 1–8 and rebinned the remaining channels into 32 bins using the *ROSAT* Standard Analysis Software System binning. We have verified that neglect of the first and last two points of the rebinned spectra does not change

**Figure 2.** The Kaz 163 0.1–2.5 keV light curve. The data have been binned into 400-s bins (corresponding to the wobble period), and are corrected for detector dead time, vignetting and wire shadowing. The count rate of Kaz 163 varies by $\sim$ 45 per cent over 55 000 s.

the nature of our fitting results below.

Systematic errors of 2 per cent were added in quadrature to the data point rms errors, to account for residual uncertainties in the spectral calibration of the PSPC. We have used the 1993 January 15 response matrix (MPE No. 36) for all analysis. This matrix corrects for the systematic deficit of photons near the PSPC detector's carbon edge that was present in earlier matrices (cf. Turner, George & Mushotzky 1993). The expected systematic errors from this matrix are a few per cent, and are significantly smaller than the relevant residuals in the data discussed below. We are keenly aware, however, of remaining spectral calibration uncertainties with the PSPC, and refer the reader to the discussions of this issue in appendix A of Brinkmann et al. (1994), and appendix A and appendix B of Fiore et al. (1994). Our results below agree with the general trend of steep soft Seyfert spectra, and the results for Ark 564 are in line with what has been seen for other narrow-line Seyfert 1s. It is important to note that the spectra of Ark 564 and other narrow-line Seyfert 1 galaxies are steep *even when compared to* those of other *ROSAT* -observed Seyfert 1 galaxies.

We model the the X-ray spectra of our targets using the X-ray spectral models in the XSPEC spectral fitting package (Shafer et al. 1991).

### 3.3.1   Ark 564

The results of basic spectral fits to the Ark 564 spectrum are shown in Table 2. The errors for these and all other fits are quoted for 68.3 per cent confidence (unless explicitly stated otherwise), taking all free parameters to be of interest other than absolute normalization (Lampton, Margon & Bowyer 1976; Press et al. 1989). A single absorbed power-law model fits the spectrum poorly, yielding $\chi^2_\nu = 2.21$ (see Fig. 3a). Large systematic residuals are clearly visible. The derived photon index of $\Gamma_1 = 3.39^{+0.07}_{-0.07}$ is significantly higher than that seen in higher energy *Ginga* observations of Seyfert galaxies. Nandra & Pounds (1994), for example, find the mean X-ray continuum of 27 Seyferts to be best-modelled by a power law with photon index in the range $\Gamma = 1.9$–2.0. A high photon index could be due to a number of effects. For example, spectral features such as edges and lines or an additional soft continuum component could mimic a steep spectrum. We have examined these possibilities by fitting models with spectral features and additional soft continuum components, and the results of our fitting are shown in Table 2.

Models with simple soft continuum components reduce $\chi^2_\nu$, but have other problems. For example, the power-law and bremsstrahlung model yields a *negative* $\Gamma_1$ (albeit with large error bars; note that photon flux is proportional to $E^{-\Gamma_1}$) as well as an unusually strong bremsstrahlung component. A two-power-law model (not shown in Table 2) yields an unphysical photon index for the secondary power law of $\Gamma_2 > 9$. The values of $N_H$ derived for both the bremsstrahlung and blackbody soft component models are too small to agree with the Galactic column density (see Table 1). For the power-law and blackbody model, the 90 per cent confidence $N_H = (3.41^{+1.90}_{-1.30}) \times 10^{20}$ cm$^{-2}$, which still does not overlap the Galactic $N_H$ value. This disagreement is only aggravated by the fact that Ark 564 may have intrinsic absorption. Walter & Fink (1993) argue for the strong absorption of the ultraviolet continuum of Ark 564 by dust, and gas associated with dust will absorb X-rays. Assuming that the X-rays and ultraviolet radiation travel through the same matter, a Galactic dust-to-gas ratio, case-

**Table 2.** Basic spectral fitting to Ark 564.

| Model Name | $N_{\rm H}$ /($10^{20}$ cm$^{-2}$) | $A_1$ /($10^{-3}$ ph keV$^{-1}$ cm$^{-2}$ s$^{-1}$) | $\Gamma_1$ | Other Parameters | $\chi^2_\nu$/d.o.f. |
|---|---|---|---|---|---|
| Power Law | $7.29^{+0.26}_{-0.26}$ | $12.00^{+0.25}_{-0.24}$ | $3.39^{+0.07}_{-0.07}$ | — | 2.21/29 |
| Power Law & Brems. | $4.84^{+0.43}_{-0.43}$ | $0.59^{+1.67}_{-0.59}$ | $-0.92^{+3.24}_{-0.03}$ | $K = (12.12^{+3.91}_{-2.19}) \times 10^{-2}$<br>$kT = 0.40^{+0.05}_{-0.05}$ keV | 1.14/27 |
| Power Law & Blackbody | $3.41^{+1.31}_{-1.13}$ | $5.85^{+2.29}_{-2.29}$ | $2.10^{+0.73}_{-0.74}$ | $K = (3.69^{+0.65}_{-1.27}) \times 10^{-4}$<br>$kT = 0.13^{+0.01}_{-0.01}$ keV | 0.93/27 |
| Power Law & Edge | $6.13^{+0.52}_{-0.51}$ | $13.72^{+0.73}_{-0.70}$ | $2.93^{+0.18}_{-0.18}$ | $\tau = 0.84^{+0.31}_{-0.30}$<br>$E_{\rm Edge} = 1.15^{+0.06}_{-0.05}$ keV | 0.85/27 |
| Power Law & Gaussian Line | $6.48^{+0.59}_{-0.63}$ | $10.83^{+0.59}_{-0.59}$ | $3.22^{+0.16}_{-0.17}$ | $K_{\rm Line} = (2.02^{+1.14}_{-0.94}) \times 10^{-3}$ cm$^{-2}$ s$^{-1}$<br>$E_{\rm Line} = 0.79^{+0.07}_{-0.06}$ keV | 1.31/27 |

All spectral models above are described in Shafer et al. (1991), and the spectral parameters are defined as follows: $N_{\rm H}$ = equivalent hydrogen column in atom cm$^{-2}$; $A_1$ = normalization of power law in photon keV$^{-1}$ cm$^{-2}$ s$^{-1}$; $\Gamma_1$ = photon index of power law; $K$ = dimensionless normalization of bremsstrahlung or blackbody; $kT$ = bremsstrahlung or blackbody temperature in keV; $\tau$ = edge absorption depth at threshold; $E_{\rm Edge}$ = edge threshold energy (corrected for redshift); $K_{\rm Line}$ = line flux; $E_{\rm Line}$ = line energy (corrected for redshift).
All error bars are for 68.3 per cent confidence, assuming that all free parameters are of interest other than the absolute normalization. The Gaussian line is taken to have a fixed width of 50 eV.

B recombination, a density of $10^4$ cm$^{-3}$ and a temperature of 20 000 K (the density and temperature chosen are characteristic for the narrow-line region and our results do not depend strongly on their values), we can estimate $E(B-V)$ (Gaskell & Ferland 1984; Ward et al. 1987; Osterbrock 1989) and then the equivalent $N_{\rm H}$ column (Bohlin, Savage & Drake 1978; Burstein & Heiles 1978) expected from Ark 564's large narrow-line Balmer decrement of 4.4 (see Section 1). We estimate an intrinsic column of $N_{\rm H} \sim 2 \times 10^{21}$ cm$^{-2}$, consistent with the prediction of Walter & Fink (1993). While our $N_{\rm H}$ estimation is admittedly crude (and probably breaks down as we argue below), any intrinsic column adds weight to the evidence against a simple soft continuum component. We discuss more complex soft continuum components below.

The power-law and edge model for the data indicates an edge energy of $1.15^{+0.06}_{-0.05}$ keV. In the PSPC band, the dominant absorber is oxygen, because of both its large abundance (Morrison & McCammon 1983) and its large photoionization cross-section (Daltabuit & Cox 1972). It has K$\alpha$ edge energies ranging from 0.533 keV (for O I) to 0.870 keV (for O VIII), and the edge energy we fit is inconsistent with this energy range. While it is possible that combined edges from neon, sodium, and iron (the L edge) could, if oxygen is fully stripped, produce the feature we detect, this interpretation is implausible for several reasons. Both neon and sodium are usually significantly less abundant than oxygen, and also have smaller photoionization cross-sections. They would therefore have difficulty producing an edge with depth $\tau = 0.84^{+0.31}_{-0.30}$. In addition, there is only a very narrow range of ionization parameters over which oxygen is fully or almost fully stripped while neon and sodium can still form edges (see, for example, fig. 12 of Kallman & McCray 1982). Iron L edges alone cannot produce an edge feature at $1.15^{+0.06}_{-0.05}$ keV if oxygen is fully stripped, since iron will be in the ionization stages of Fe XVIII (which has an L edge energy of 1.358 keV) and above. Furthermore, it is worth noting that possible absorption by outflowing material (which would raise the edge energy) predicts an extremely large outflow velocity, which would cause kinetic outflow power to overwhelm radiation power (unless the outflow were *extremely* collimated) and be in discord with the accretion hypothesis. The warm absorber model of Yaqoob & Warwick (1991) is rejected at > 96 per cent confidence using the F-test (Bevington 1969). Similarly, a warm absorber model constructed using the photoionization code CLOUDY (Ferland 1992; Fabian et al. 1994) is rejected with > 99 per cent confidence (the model differs from that of Yaqoob & Warwick 1991 in that it includes spectral line emission).

A power-law and Gaussian line model fits the data reasonably well and is physically acceptable. The line energy of $0.79^{+0.07}_{-0.06}$ keV lies in the centre of the strong O VIII K$\alpha$, Fe XVII L$\alpha$ and Fe XVIII L$\alpha$ line complex, and these lines would be mixed together by the limited PSPC energy response. They could be formed in a photoionized disc (see, for example, fig. 2 of Ross & Fabian 1993 and fig. 6a of Życki et al. 1994). The value of $N_{\rm H}$ is consistent with the Galactic value.

More complex soft continuum models with, for example, a two-blackbody soft excess can also fit the spectrum well and are entirely plausible. Due to the limited number ($\approx 5$) of independent PSPC energy bands, we are not able to probe such models in detail.

### 3.3.2  NGC 985

NGC 985 is also poorly fitted by a simple power law (see Fig 3b), and we show the results of fitting more complex models in Table 3. A power-law and line model does not describe these data well, as it predicts a strong line at $0.46^{+0.06}_{-0.06}$ keV. While the C VI K$\alpha$ line energy is marginally consistent with this energy, it is hard to imagine a plausible scenario where it is formed with such a high intensity without the stronger

**Figure 3.** Spectra for (a) Ark 564 and (b) NGC 985 in the 0.1–2.5 keV band. Single power-law fits to the observations and residuals are also shown. Note that neither spectrum is well fitted by a simple power-law model.

O K$\alpha$ and Fe L$\alpha$ lines also being formed (which would then lead to a higher fit energy). In addition, the derived $N_{\rm H}$ is low, being only barely consistent with the Galactic value (see Table 1 and note that the 90 per cent confidence $N_{\rm H} = (2.21^{+0.69}_{-0.51}) \times 10^{20}$ cm$^{-2}$).

Models with soft continuum components do fit NGC 985's spectrum well and give physically plausible results. The $N_{\rm H}$ for the power-law and blackbody model agrees very well with the Galactic value, and the power-law and bremsstrahlung model $N_{\rm H}$ is feasible if there is intrinsic absorption. The soft components are powerful, dominating the spectrum up to about 0.6 keV. A power-law and edge model also fits the spectrum of NGC 985 well and gives a reasonable edge energy. A natural mechanism for the formation of such an edge is oxygen absorption in a warm absorber, and the warm absorber model of Yaqoob & Warwick (1991) fits the data with an ionization parameter of $U = 1.12^{+0.24}_{-0.23}$, a warm absorber $N_{\rm H}$ of $(6.67^{+2.93}_{-2.21}) \times 10^{21}$ cm$^{-2}$, a power-law photon index of $2.73^{+0.18}_{-0.19}$, and $\chi^2_\nu = 0.64$. The photon index derived in the warm absorption scenario is again high, hinting that, if NGC 985 has a warm absorber, then it probably has a continuum soft excess as well.

### 3.3.3  Kaz 163, Mrk 79 and RX J2256.6+0525

In Table 4 we show the results of single power-law fits to Kaz 163, Mrk 79 and RX J2256.6+0525. For the fixed photon index fits, we use the Nandra & Pounds (1994) mean intrinsic photon index of 1.95. While the three fits with variable spectral indices yield formally acceptable values of $\chi^2_\nu$, their steep spectra belie this simple interpretation. All three are inconsistent with the Nandra & Pounds (1994) $\Gamma_{\rm int} = 1.95^{+0.05}_{-0.05}$ (and are even less consistent with single absorbed power-law fits to *HEAO-1*, *EXOSAT*, and *Ginga* data; see Nandra & Pounds 1994, fig. 2). The derived $N_{\rm H}$ for Kaz 163 is low, and is only marginally consistent with the Galactic value (note that the 90 per cent confidence $N_{\rm H} = (3.73^{+0.42}_{-0.41}) \times 10^{20}$ cm$^{-2}$). Fits with a fixed power-law photon index and extra components do significantly improve the $\chi^2_\nu$s over those for fixed photon indices in Table 4. Due to our assumption that $\Gamma_1 = 1.95$, a detailed presentation of the spectral modelling of these sources would not be appropriate. We do, however, report the best secondary component fits so as to probe the spectra to the greatest extent possible. Kaz 163 is best fitted with a secondary power-law component. It gives $N_{\rm H} = (5.01^{+1.40}_{-1.27}) \times 10^{20}$ cm$^{-2}$ (removing the $N_{\rm H}$ disagreement), $\Gamma_2 = 3.59^{+0.81}_{-1.03}$, and $\chi^2_\nu = 0.85$ for 28 degrees of freedom. Mrk 79 is best fitted with a $0.21^{+0.10}_{-0.05}$ keV secondary bremsstrahlung component with $N_{\rm H} = (8.00^{+1.34}_{-1.11}) \times 10^{20}$ cm$^{-2}$ and $\chi^2_\nu = 0.61$ for 28 degrees of freedom. RX J2256.6+0525 is best fitted with a blackbody secondary component. The best-fitting parameters are $kT = 0.19^{+0.05}_{-0.05}$ keV, $N_{\rm H} = (5.55^{+1.00}_{-1.15}) \times 10^{20}$ cm$^{-2}$, and $\chi^2_\nu = 0.83$ for 28 degrees of freedom.

## 4  DISCUSSION AND CONCLUSIONS

The geometry, composition and ionization state of matter around Seyfert 1 engines are of central importance when trying to understand Seyfert 1 spectra. We have analysed pointed *ROSAT* observations of five Seyfert galaxies, and find evidence in all five for complex soft emission. This emission may be modelled in our various Seyfert 1s as O VIII K$\alpha$, Fe XVII L$\alpha$, and Fe XVIII L$\alpha$ line emission, powerful

**Table 3.** Basic spectral fitting to NGC 985.

| Model Name | $N_H$ /$(10^{20}$ cm$^{-2})$ | $A_1$ /$(10^{-3}$ ph keV$^{-1}$ cm$^{-2}$ s$^{-1})$ | $\Gamma_1$ | Other Parameters | $\chi^2_\nu$/d.o.f. |
|---|---|---|---|---|---|
| Power Law | $3.09^{+0.26}_{-0.25}$ | $3.64^{+0.10}_{-0.10}$ | $2.58^{+0.10}_{-0.09}$ | — | 2.57/29 |
| Power Law & Brems. | $4.17^{+1.10}_{-0.85}$ | $2.80^{+0.55}_{-0.79}$ | $1.62^{+0.45}_{-0.64}$ | $K = (14.5^{+14.20}_{-8.05}) \times 10^{-2}$<br>$kT = 0.18^{+0.07}_{-0.05}$ keV | 0.98/27 |
| Power Law & Blackbody | $3.00^{+0.98}_{-0.74}$ | $2.99^{+0.36}_{-0.41}$ | $1.75^{+0.33}_{-0.38}$ | $K = (1.74^{+1.20}_{-0.51}) \times 10^{-4}$<br>$kT = 0.08^{+0.01}_{-0.01}$ keV | 0.80/27 |
| Power Law & Edge | $3.57^{+0.53}_{-0.47}$ | $4.76^{+0.40}_{-0.38}$ | $2.57^{+0.17}_{-0.16}$ | $\tau = 0.95^{+0.31}_{-0.29}$<br>$E_{\rm Edge} = 0.76^{+0.06}_{-0.05}$ keV | 0.63/27 |
| Power Law & Gaussian Line | $2.21^{+0.53}_{-0.41}$ | $3.41^{+0.18}_{-0.16}$ | $2.25^{+0.17}_{-0.16}$ | $K_{\rm Line} = (5.03^{+3.94}_{-2.24}) \times 10^{-3}$ cm$^{-2}$ s$^{-1}$<br>$E_{\rm Line} = 0.46^{+0.06}_{-0.06}$ keV | 1.37/27 |

All spectral models above are described in Shafer et al. (1991), and the spectral parameters are defined as follows: $N_H$ = equivalent hydrogen column in atom cm$^{-2}$; $A_1$ = normalization of power law in photon keV$^{-1}$ cm$^{-2}$ s$^{-1}$; $\Gamma_1$ = photon index of power law; $K$ = dimensionless normalization of bremsstrahlung or blackbody; $kT$ = bremsstrahlung or blackbody temperature in keV; $\tau$ = edge absorption depth at threshold; $E_{\rm Edge}$ = edge threshold energy (corrected for redshift); $K_{\rm Line}$ = line flux; $E_{\rm Line}$ = line energy (corrected for redshift).
All error bars are for 68.3 per cent confidence, assuming that all free parameters are of interest other than the absolute normalization. The Gaussian line is taken to have a fixed width of 50 eV.

soft continuum emission, and intrinsic warm absorption.

The combination of unusual observational parameters for Ark 564 makes it a challenge to model. We will first summarize Ark 564's features, and will then examine the explanatory extent and feasibility of various models. Basic spectral fitting of Ark 564 consistently yields very steep ($\Gamma_1 \approx 3$) spectra for this powerful (its 0.2–2.0 keV luminosity is $\sim 2.4 \times 10^{44}$ erg s$^{-1}$) narrow-line Seyfert 1. This is in line with what has been seen in other narrow-line Seyfert 1 galaxies (see Section 1). In the absence of flattening at energies below 0.1 keV, extrapolated soft X-ray spectra overpredict the dereddened (Savage & Mathis 1979) 2675-Å and 1375-Å fluxes (Walter & Fink 1993) by factors $> 160$ and $> 120$, respectively. A naive estimation (see Section 3.3.1 above) of Ark 564's intrinsic column suggests a possible $N_H$ of up to $2 \times 10^{21}$ cm$^{-2}$ from gas associated with the dust creating its narrow-line Balmer decrement (see Section 3.3), yet we see negligible X-ray absorption over the Galactic column. Ark 564 has strong optical Fe II lines (perhaps not surprising in light of its steep X-ray spectrum and fig. 2 of Wilkes, Elvis & McHardy 1987). The line emission from its broad-line region is relatively narrow for a Seyfert 1 galaxy (mirabile dictu, its classification as a narrow-line Seyfert 1 galaxy), indicating that the line-emitting clouds have lower velocities along the line of sight than do typical Seyfert 1 galaxies. Ark 564's 0.1–2.5 keV flux varies by $\sim 20$ per cent in 1 500 s.

From the large $N_H$ disparity, we conclude that one of the assumptions used in the conversion from Balmer decrement to $N_H$ must break down (see Section 3.3.1). One possibility is that the assumption that a Galactic dust-to-gas ratio is appropriate for Ark 564 is invalid. This possibility has been explored by Walter & Fink (1993). If true, it indicates that Ark 564 has a dust-to-gas ratio of at least $\sim 20$ times the Galactic value. This high value, while not impossible, is somewhat surprising in light of the fact that the amount of dust will be reduced by thermal sputtering, photodestruction at cloud surfaces, and shocks from high-speed cloud collisions (Krolik & Begelman 1988; section 7 of Laor & Draine 1993). This model does not make any attempt to connect the generally steep soft X-ray spectra being seen in narrow-line Seyfert 1 galaxies with the fact that they have slimmer optical emission lines than 'normal' Seyfert 1s.

The assumption that the X-rays and ultraviolet emission travel through the same matter might also break down. This certainly seems likely at least to some degree in the inhomogeneous neighborhood thought to exist around a Seyfert 1 nucleus. Gas and dust near the central engine will be made inhomogeneous as they are photodestroyed, photoionized, clumped, and blown around by anisotropic radiation pressure and supersonic winds (Wolfe 1974). X-rays emitted near the central engine might, for example, escape relatively unabsorbed down an 'ionization cone' that has its apex at the nucleus and extends beyond the classical narrow-line region (e.g. Unger et al. 1987; Wilson, Ward & Haniff 1988; Acosta-Pulido et al. 1990). The more generally extended ultraviolet emission will suffer more more absorption and will be weakened relative to the X-rays. However, this model cannot easily explain the lack of very broad lines without invoking a somewhat ad hoc arrangement of dust and gas around the broad-line region. Potential radiation scattering models are limited by the fact that Ark 564 shows no evidence for polarization (Goodrich 1989). This lack of polarization also hints that Ark 564 has only a weakened inner broad-line region, if any, where broad-line clouds would move most rapidly (as opposed to a hidden broad-line region). This model also does not establish a connection between the steep X-ray slope and relatively slim optical emission lines.

The assumption of case-B recombination might fail for Ark 564, due to its powerful soft X-ray flux. A strong soft X-ray component will increase the amount of X-ray heating of

**Table 4.** Power-law fits to Kaz 163, Mrk 79 and RX J2256.6 + 0525.

| Seyfert Name | $N_H$ /($10^{20}$ cm$^{-2}$) | $A_1$ /($10^{-3}$ ph keV$^{-1}$ cm$^{-2}$ s$^{-1}$) | $\Gamma_1$ | $\chi^2_\nu$/d.o.f. |
|---|---|---|---|---|
| Kaz 163 | $3.73^{+0.30}_{-0.29}$ | $0.94^{+0.03}_{-0.03}$ | $2.54^{+0.10}_{-0.10}$ | 0.92/29 |
|  | $2.08^{+0.05}_{-0.05}$ | $0.90^{+0.02}_{-0.02}$ | 1.95 Fixed | 4.02/30 |
| Mrk 79 | $7.53^{+0.46}_{-0.45}$ | $10.30^{+0.30}_{-0.29}$ | $2.57^{+0.11}_{-0.11}$ | 0.84/29 |
|  | $5.20^{+0.11}_{-0.11}$ | $9.48^{+0.17}_{-0.16}$ | 1.95 Fixed | 3.63/30 |
| RX J2256.6 + 0525 | $8.08^{+0.98}_{-0.88}$ | $1.67^{+0.09}_{-0.09}$ | $2.48^{+0.18}_{-0.18}$ | 0.85/29 |
|  | $5.87^{+0.27}_{-0.24}$ | $1.54^{+0.05}_{-0.05}$ | 1.95 Fixed | 1.55/30 |

All spectral models above are described in Shafer et al. (1991), and the spectral parameters are defined as follows:
$N_H$ = equivalent hydrogen column in atom cm$^{-2}$; $A_1$ = normalization of power law in photon keV$^{-1}$ cm$^{-2}$ s$^{-1}$; $\Gamma_1$ = photon index of power law.
All error bars are for 68.3 per cent confidence, assuming that all free parameters are of interest other than the absolute normalization. The word "fixed" by a parameter means that it was held fixed during the fit.

neutral gas and move the Balmer decrement away from the standard case-B value toward a value more appropriate for pure collisional excitation (Gaskell & Ferland 1984). This solution, unlike the models mentioned above, can explain the $N_H$ discrepancy without requiring a peculiar dust-to-gas ratio or special arrangement of matter. The steep soft X-ray spectrum might also be *directly connected* with the lack of broad optical lines in narrow-line Seyfert 1 galaxies; a large number of soft X-ray photons might interfere with broad-line cloud confinement and formation near the central source where cloud velocities would be largest. In a simple pressure-confined broad-line cloud model, for example, a steep soft X-ray spectrum removes the double-valued behavior of the curve of gas temperature as a function of ionization parameter $\Xi$ (see fig. 1 of Guilbert, Fabian & McCray 1983 and the associated discussion) and thereby eliminates the thermal instability associated with multiphase cloud confinement. Clouds may also have difficulty forming via small perturbations in an environment without the thermal instability. This would hinder broad line formation in models where many clouds are born and die over the time required for an emission-line cloud to cross the broad-line region. While a simple pressure-confinement model is almost certainly inappropriate, generally similar behaviour might be expected in models with more complex cloud confinement (e.g. magnetically aided pressure confinement, evaporation-aided pressure confinement; Kallman & Mushotzky 1985) and cloud formation (e.g. shock-induced cloud formation; Perry & Dyson 1985) mechanisms. For example, a steep X-ray spectrum increases the difference of $\Xi$ between hot and cool states (see fig. 1 of Guilbert et al. 1983), thereby increasing the shock velocity required to form cool clouds from shocked hot gas. A steep spectrum would not affect purely magnetic confinement.

In many broad-line region models, optical Fe II emission is created when X-rays penetrate deeply into broad-line clouds and create a warm, partially ionized zone at high optical depths (Kwan & Krolik 1981). The strong Fe II lines of Ark 564 can still be plausibly formed even in the absence of broad-line clouds near the central engine by an irradiated or mechanically heated outer accretion disc (Collin-Souffrin & Dumont 1990; Joly 1987) or inner torus. O VIII K$\alpha$, Fe XVII L$\alpha$ and Fe XVIII L$\alpha$ lines could be formed in the inner radiation-dominated part of the same disc. To make the *ROSAT* and ultraviolet observations compatible, the spectrum must break below 0.1 keV. We have found that simple models with, for example, a two-blackbody soft excess can produce the required ultraviolet/X-ray spectrum. Such spectra are, of course, poorly constrained by our data but are plausible. A break at less than 0.1 keV does not reinstate the thermal instability mentioned above, as we have verified with CLOUDY (Ferland 1992) computations.

Another possibility based on orientation effects is that Ark 564 and other similar narrow-line Seyfert 1 galaxies might be those Seyfert 1s where we view the disc in a relatively 'face-on' manner. If the general flow of matter in the vicinity of the disc were largely perpendicular to the line of sight, this would explain narrower optical lines. Ark 564's steep X-ray slope might also be understood via orientation, due to the angularly dependent soft X-ray flux and reflectivity of an accretion disc (see fig. 3 of Madau 1988, fig. 2 of Sun & Malkan 1989, fig. 10 of Zycki et al. 1994 and the associated discussions). Statistical analysis of a large number of Seyfert 1s might be used to probe this possibility.

Ark 564's rôle as a bright X-ray Seyfert 1 with narrow optical emission lines (Osterbrock & Shuder 1982) can be compared with the recent work of Boyle et al. (1994) which suggests that narrow-line X-ray galaxies (NLXGs) may be a significant contributor to the soft X-ray background. Ark 564's Balmer decrement is consistent with the large Balmer decrements seen in NLXGs, and the NLXGs of Boyle et al. (1994) also show little evidence for $N_H$ values in excess of the Galactic value despite their large Balmer decrements (B.J. Boyle, private communication). Ark 564's place as a potential low-redshift counterpart to these objects is questionable, however, due to the fact that its steep soft X-ray spectrum is not generally observed and the fact that its [O III] to H$\beta$ ratio of 12.5 (Stirpe 1990) is much larger than is generally seen in these objects.

We do not detect reprocessed X-ray line emission in NGC 985 from the postulated accretion disc of Stanga et al. (1991), but do find evidence for a powerful soft X-ray excess (which may arise via disc reprocessing and reflection) and/or a warm absorber oxygen edge. Kaz 163, Mrk 79 and RX J2256.6 + 0525 all have steep 0.1–2.5 keV spectra that are best modelled with continuum soft excess components. The variety of soft component shapes seen in these objects presages the rich physics that will be further accessible to soft X-ray detectors of higher energy resolution.

**ACKNOWLEDGMENTS**

We thank Steve Allen, Thomas Boller, Brian Boyle, Carolin Crawford, Chris Done, Alastair Edge, Liz Puchnarewicz and Martin Ward for useful discussions, and Tahir Yaqoob for providing the warm absorber model. We thank Thomas Boller for providing us with the X-ray data for Kaz 163 and Carolin Crawford for providing us with the optical spectrum of RX J2256.6+0525. We thank R. Mushotzky for his kind help with the *HEAO-1* data base. We gratefully acknowledge help from members of the Institute of Astronomy X-ray group, and financial support from the United States National Science Foundation and the British Overseas Research Studentship Programme (WNB), the PPARC (KN, CSR) and the Royal Society (ACF). The *ROSAT* project is supported by the Bundesministerium für Forschung und Technologie (BMFT). Much of our analysis has relied on the Starlink ASTERIX X-ray data processing system and the XSPEC X-ray spectral fitting software, and we thank the people who have created and maintain this software. This research has made use of data obtained from the UK *ROSAT* Data Archive Centre at the Department of Physics and Astronomy, University of Leicester and has also made use of the NASA/IPAC extragalactic data base (Helou et al. 1991) which is operated by the Jet Propulsion Laboratory, Caltech.

**REFERENCES**

Acosta-Pulido J.A., Pérez-Fournon I., Calvani M., Wilson A.S., 1990, ApJ, 365, 119
Arnaud M., Lachieze-Rey M., Rothenflug R., Yamashita K., Hatsukade I., 1991, A&A, 243, 56
Bevington P., 1969, Data Reduction and Error Analysis for the Physical Sciences. McGraw Hill, New York
Bohlin R.C., Savage B.D., Drake J.F., 1978, ApJ, 224, 132
Boller T., Trümper J., Molendi S., Fink H., Schaeidt S., Caulet A., Dennefeld M., 1993, A&A, 279, 53
Boyle B.J., McMahon R.G., Wilkes B.J., Elvis M., 1994, MNRAS, *submitted*
Brinkmann W. et al., 1994, A&A, *submitted*
Burstein D., Heiles C., 1978, ApJ, 225, 40
Collin-Souffrin S., Dumont A.M., 1990, A&A, 229, 292
Crawford C.S., Edge A.C., Fabian A.C., Allen S.W., Böhringer H., Ebeling H., McMahon R.G., Voges W., 1994, MNRAS, *submitted*
Daltabuit E., Cox D.P., 1972, ApJ, 177, 855
Elvis M., Lockman F.J., Wilkes B.J., 1989, AJ, 97, 777
Fabian A.C., 1992, in Holt S.S., Neff S.G., Urry C.M., eds, Testing the AGN Paradigm. AIP Press, New York, p. 657
Fabian A.C. et al., 1994, PASJ, 46, L59
Ferland G.J., 1992, Univ. of Kentucky Department of Physics and Astronomy Internal Report
Fiore F., Elvis M., McDowell J.C., Siemiginowska A., Wilkes B.J., 1994, ApJ, *submitted*
Gaskell C.M., Ferland G.J., 1984, PASP, 96, 393
Goodrich R.W., 1989, ApJ, 342, 224
Gondhalekar P.M., Kellett B.J., Pounds K.A., Matthews L., Quenby J.J., 1994, MNRAS, 268, 973
Guilbert P.W., Fabian A.C., McCray R., 1983, ApJ, 266, 466
Hasinger G., Turner T.J., George I.M., Boese G., 1992, Legacy #2, The Journal of the High Energy Astrophysics Science Archive Research Center, NASA/GSFC
Helou G., Madore B.F., Schmitz M., Bicay M.D., Wu X., Bennett J., 1991, in Egret D., Albrecht M., eds, Databases and On-Line Data in Astronomy, Kluwer, Dordrecht, p. 89
Hutchings J.B., Hickson P., 1988, AJ, 95, 1363
Johnson M.W., Cruddace R.G., Ulmer M.P., Kowalski M.P., Wood K.S., 1983, ApJ, 266, 425
Joly M., 1987, A&A, 184, 33
Kallman T., McCray R., 1982, ApJS, 50, 263
Kallman T., Mushotzky R., 1985, ApJ, 292, 49
Kriss G.A., Canizares C.R., 1982, ApJ, 261, 51
Krolik J.H., Begelman M.C., 1988, ApJ, 329, 702
Kruper J.S., Urry C.M., Canizares C.R., 1990, ApJS, 74, 347
Kwan J., Krolik J.H., 1981, ApJ, 250, 478
Lampton M., Margon B., Bowyer S., 1976, ApJ, 208, 177
Laor A., Draine B.T., 1993, ApJ, 402, 441
Madau P., 1988, ApJ, 327, 116
Mazzarella J.M., Boroson T.A., 1993, ApJS, 85, 27
Morrison R., McCammon D., 1983, ApJ, 270, 119
Mushotzky R.F., Done C., Pounds K.A., 1993, ARA&A, 31, 717
Nandra K., Pounds K.A., 1994, MNRAS, 268, 405
Nandra K. et al., 1993, MNRAS, 260, 504
Oke J.B., Lauer T.R., 1979, ApJ, 230, 360
Osterbrock D.E., 1989, Astrophysics of Gaseous Nebulae and Active Galactic Nuclei. Univ. Science Books, Mill Valley
Osterbrock D.E., Pogge R.W., 1985, ApJ, 297, 166
Osterbrock D.E., Shuder J.M., 1982, ApJS, 49, 149
Perry J.J., Dyson J.E., 1985, MNRAS, 213, 665
Pfeffermann E. et al., 1987, Proc. SPIE, 733, 519
Press W.H., Flannery B.P., Teukolsky S.A., Vetterling W.T., 1989, Numerical Recipes in Pascal. Cambridge Univ. Press, Cambridge
Puchnarewicz E.M., Mason K.O., Córdova F.A., Kartje J., Branduardi-Raymont G., Mittaz J.P.D., Murdin P.G., Allington-Smith J., 1992, MNRAS, 256, 589
Rodríguez Espinosa J.M., Stanga R.M., 1990, ApJ, 365, 502
Ross R.R., Fabian A.C., 1993, MNRAS, 261, 74
Savage B.D., Mathis J.S., 1979, ARA&A, 17, 73
Shafer R.A., Haberl F., Arnaud K.A., Tennant A.F., 1991, XSPEC Users Guide. ESA Publications, Noordwijk
Snowden S., Plucinsky P.P., Briel U., Hasinger G., Pfeffermann E., 1992, ApJ, 393, 819
Stanga R.M., Rodríguez Espinosa J.M., Mannucci F., 1991, ApJ, 379, 592
Stark A.A., Gammie C.F., Wilson R.W., Bally J., Linke R., Heiles C., Hurwitz M., 1992, ApJS, 79, 77
Stirpe G.M., 1990, A&AS, 85, 1049
Stocke J.T., Morris S.L., Gioia I.M., Maccacaro T., Schild R., Wolter A., Fleming T.A., Henry J.P., 1991, ApJS, 76, 813
Stone O., 1886, AJ, 7, 57
Sun W., Malkan M.A., 1989, ApJ, 346, 68
Trümper J., 1983, Adv. Space Res., 4, 241
Turner T.J., Pounds K.A., 1989, MNRAS, 240, 833
Turner T.J., George I.M., Mushotzky R.F., 1993, ApJ, 412, 72
Ulvestad J.S., Wilson A.S., 1984, ApJ, 278, 544
Unger S.W., Pedlar A., Axon D.J., Whittle M., Meurs E.J.A., Ward M.J., 1987, MNRAS, 228, 671
van Groningen E., 1993, A&A, 272, 25
Walter R., Fink H.H., 1993, A&A, 274, 105

Ward M.J., Geballe T., Smith M., Wade R., Williams P., 1987, ApJ, 316, 138
Wilkes B.J., Elvis M., McHardy I., 1987, ApJ, 321, L23
Wilson A.S., Ward M., Haniff C.A., 1988, ApJ, 334, 121
Wolfe A.M., 1974, ApJ, 188, 243
Yaqoob T., Warwick R.S., 1991, MNRAS, 248, 773
Życki P.T., Krolik J.H., Zdziarski A.A., Kallman T.R., 1994, ApJ, *submitted*

This paper has been produced using the Blackwell Scientific Publications TeX macros.